\documentclass[floatfix,aps,showpacs,twocolumn,nofootinbib,preprintnumbers]
{revtex4}
\usepackage{graphicx}
\usepackage{epsfig}
\usepackage{bm}
\usepackage{longtable}

\def\lsim{~\rlap{$<$}{\lower 1.0ex\hbox{$\sim$}}\;}
\def\gsim{~\rlap{$>$}{\lower 1.0ex\hbox{$\sim$}}\;}

\begin{document}

\preprint{IFT-P.038/2006}

\title{Cosmic microwave background and large-scale structure 
constraints on a simple quintessential inflation model}
\author{Rogerio~Rosenfeld} \email{rosenfel@ift.unesp.br}
\affiliation{Instituto de F\'{\i}sica Te\'orica - UNESP, Rua
Pamplona, 145, 01405-900, S\~{a}o Paulo, SP, Brazil}
\author{Joshua~A.~Frieman} \email{frieman@fnal.gov}
\affiliation{Particle Astrophysics Center, Fermi
        National Accelerator Laboratory, Batavia, Illinois \ 60510-0500, USA,\\
        Department of Astronomy and Astrophysics, Kavli Institute for Cosmological Physics,
        University of Chicago, Chicago, Illinois \ 60637-1433, USA}

\date{\today}

\begin{abstract}

We derive constraints on a simple quintessential inflation
model, based on a spontaneously broken $\Phi^4$ theory, imposed by
the Wilkinson Microwave Anisotropy Probe three-year 
data (WMAP3) and by galaxy clustering results from the Sloan Digital Sky Survey (SDSS). 
We find that the scale of symmetry breaking must be larger than about 3 Planck masses in order 
for inflation to generate acceptable values of the scalar spectral index and of 
the tensor-to-scalar ratio. We also show that the resulting quintessence 
equation-of-state can evolve rapidly at recent times and hence can potentially be 
distinguished from a simple cosmological constant in this parameter regime.

\end{abstract}


\maketitle

\section{Introduction}

The inflationary scenario, in which the Universe undergoes a phase of accelerated 
expansion in its very early moments, provides an attractive solution 
to the flatness and horizon puzzles of the standard Big Bang cosmology. In addition, 
via quantum fluctuations, it naturally provides the seed perturbations 
for later structure formation in the Universe \cite{LindeBook}. After inflation, 
reheating leads to a radiation-dominated era, followed by 
a more recent epoch in which the density is dominated by non-relativistic 
matter. 

However, there is now solid observational evidence from Type Ia supernovae
(SNIa) that the Universe is undergoing another  
burst of accelerated expansion; in the context of General Relativity, 
this must be fuelled by an unknown component
with negative pressure, usually called dark energy (DE)
\cite{Reviewdarkenergy}. 

The simplest possibility for the DE is the cosmological constant, $\Lambda$. 
Data from the cosmic microwave background (CMB) \cite{wmap}, large-scale 
structure \cite{lss}, and SNIa \cite{SNIaRiess,SNIaLegacy} are all 
consistent with the $\Lambda$CDM model, a nearly flat
Universe with a cosmological constant and nearly scale-invariant 
primordial perturbations. In the best-fit $\Lambda$CDM model, 
the vacuum energy makes up 74$\%$ of
the critical density, and the remainder is non-relativistic cold
dark matter (CDM, 22$\%$) and baryonic matter (4$\%$).

While the cosmological constant is compatible with the current data, the recognition 
that the Universe appears to have undergone more than one period of accelerated 
expansion points to the plausibility of alternative explanations for the 
dark energy. In fact, there are no consensus particle physics models for 
either primordial inflation or the recent acceleration of the Universe. 
A first step out of our ignorance is often taken by
introducing simple models, usually involving scalar fields. We
can then proceed to test these models against data and obtain
constraints on parameters that we hope can be calculated from a
more fundamental theory. In this spirit, inflation and dark energy
are often modelled via scalar fields, called the inflaton and the
quintessence field. 

In general, these two fields are treated as totally independent. In
ref.~\cite{us}, we introduced a simple, well-motivated model that
unifies these two fields into a single complex scalar field. We briefly
review this model below.

We start with the general, renormalizable Lagrangian describing a complex 
scalar field $\Phi$; with appropriate choice of coupling constants, the associated 
global $U(1)$ symmetry is spontaneously broken at a high
energy scale $f$ \cite{axions}. The broken symmetry generates a flat potential
for the phase of the complex field, $\varphi$, which at this stage is a massless 
Nambu-Goldstone boson. At a much lower energy scale, $M<<f$, 
instanton or other effects explicitly break the residual symmetry, 
providing a small mass for $\varphi$, now called a 
pseudo-Nambu Goldstone boson (PNGB). The QCD axion, a by-product of a solution 
to the strong CP problem, is an example of this phenomenon. PNGBs 
in the more general context are also sometimes called axions, a usage 
into which we shall lapse, but 
we emphasize that we are not here considering the QCD axion.

The resulting low-energy effective Lagrangian is given by:
\begin{equation} \label{eq:lagrangian1}
{\cal L} =  \partial_\mu \Phi \partial^\mu \Phi^\ast - V(\Phi) +
M^4 [ \cos(\mbox{Arg} (\Phi)) -1 ]~,
\end{equation}
with the renormalizable potential
\begin{equation}
V(\Phi) = \lambda \left( \Phi \Phi^\ast - \frac{f^2}{2}\right)^2.
\label{potential}
\end{equation}

Writing the complex field $\Phi$ as
\begin{equation}
\Phi = \frac{1}{\sqrt{2}} \phi e^{i \varphi/f}
\end{equation}
we identify the modular ($\phi$) and phase ($\varphi$) parts of
$\Phi$ with the inflaton and the quintessence fields. 
A model in which the quintessence is such an axion-like PNGB field
was introduced by Frieman {\it et al.} \cite{friemanetal}. The model 
contains two mass scales, $f$ and $M$, and one dimensionless coupling 
constant, $\lambda$. As in the QCD case, we imagine that the lower 
scale $M$ is generated dynamically, by non-perturbative  
effects, leaving $f$ as the only fundamental mass scale in the theory.
As is usually done, we have set the cosmological constant to zero. 

In order for $\varphi$ to serve as dark energy, it must have 
not become dynamical until recently; otherwise, it would now be oscillating 
on a timescale short compared to the current Hubble time and would 
act instead as non-relativistic dark matter \cite{friemanetal}.
Therefore we require
\begin{equation}
m_{\varphi}  = \frac{M^2}{f} \lesssim 3 H_0
\end{equation}
where $H_0 = 100\; h$ km/s/Mpc is the Hubble parameter today. On
the other hand, for the energy density in the $\varphi$ field to
have the correct order of magnitude to explain the acceleration, 
we must have
\begin{equation}
M^4 \simeq \rho^{(0)}_c = \frac{3 H_0^2 M_{Pl}^2}{8 \pi}.
\label{Mrho}
\end{equation}
Combining these two requirements results in \cite{friemanetal}:
\begin{equation}
f > \frac{M_{Pl}}{\sqrt{24 \pi}}; \;\;\;\; M \simeq 3 \times
10^{-3} \; h^{1/2} \; \mbox{eV},
\end{equation}
where the Planck mass $M_{Pl} = 1.2 \times 10^{19}$ GeV.

This model was implemented in a hybrid inflation context by
Mass\'o and Zsembinszki \cite{MZ}. A model similar to ours, in which 
the modulus field $\phi$ is responsible for inflation and the
phase $\varphi$ produces dark matter, was studied in ref.~\cite{KSY}. In
models in which the axion-like field is the dark matter, a high energy scale for
the axion decay constant $f$ is also necessary, in order to
suppress isocurvature fluctuations to acceptable levels. In our
case, however, the axion field only becomes dynamical at such late 
times that the bounds from isocurvature fluctuations do not apply 
 \cite{ByrnesWands}.

\section{WMAP+SDSS constraints}

The 3-year data set released by the WMAP collaboration (WMAP3)
\cite{wmap} has been used by a number of authors to constrain models of
inflation \cite{AL,KKMR,SFK,Cardoso}. Of special interest to us
are the reported limits on the scalar spectral index, $n_s$, and the
ratio of tensor to scalar perturbations, $r$. Using WMAP3 plus the
large-scale power spectrum of Luminous Red Galaxies (LRGs) in the 
Sloan Digital Sky Survey (SDSS), Tegmark, et al. \cite{Tegmarketal} derive the marginalized 
constraints:
\begin{equation}
n_s = 0.967^{+0.022}_{-0.020} \; ; \;\;\; r <0.33 \; (@\; 95 \% \;
\mbox{CL}).
\end{equation}
Note that the above constraints were obtained in the context of the 
$\Lambda$CDM model, i.e., assuming that the dark energy equation of 
state is $w=-1$, and also assuming spatial flatness, massless neutrinos, 
and no running of the scalar spectral index with spatial wavelength, 
but allowing for non-zero tensor perturbations. 
Dropping one or more of those assumptions would weaken the constraints.

As shown in Fig. 19 of \cite{Tegmarketal} (reproduced below in Fig. 1), 
under the assumptions above, a simple chaotic $\lambda \phi^4$
inflationary model is marginally excluded at 95 \% CL by the WMAP3+SDSS constraints. 
Since our proposed inflation model approaches a $\lambda \phi^4$ potential 
at large values of $\phi$, one might worry that it is also disfavored by 
current data. We will see below that this is not the case in general; 
rather, values of the fundamental mass parameter $f$ below a certain level are  
excluded.

In this model, inflation is driven by the modulus field $\phi$, since the 
potential energy associated with $\varphi$ is smaller by $\sim 112$ orders of 
magnitude.
The potential $V(\Phi)$ in Eq.(\ref{potential}) in fact depends
only on $\phi$, 
\begin{equation}
V(\phi) = \frac{\lambda}{4} \left( \phi^2 \ - f^2\right)^2.
\label{vphi}
\end{equation}
Working in the context of the slow-roll approximation,
where the field evolution is slow ($\ddot\phi \simeq 0$, $\dot
\phi \simeq -V'/(3 H)$), we can define the usual slow-roll
parameters $\epsilon$ and $\eta$ \cite{inflation}:
\begin{eqnarray}
\epsilon\left(\phi\right) &=& {M_{Pl}^2 \over 16 \pi}
\left({V'\left(\phi\right)
\over V\left(\phi\right)}\right)^2; \\
\eta\left(\phi\right) &=&  {M_{Pl}^2 \over 8 \pi}
\left[{V''\left(\phi\right) \over V\left(\phi\right)} - {1 \over
2} \left({V'\left(\phi\right) \over
V\left(\phi\right)}\right)^2\right].
\end{eqnarray}
Slow roll is a consistent approximation for $V',\ V'' \ll V$ (in Planck 
units) or equivalently for $\epsilon,\ \eta \ll 1$. In particular, inflation
ends when $\epsilon \simeq 1$.

In our case we find
\begin{eqnarray}
\epsilon\left(\phi\right) &=& {M_{Pl}^2 \over \pi} {\phi^2 \over
\left(\phi^2 - f^2 \right)^2}; \label{eps}\\
\eta\left(\phi\right) &=& {M_{Pl}^2 \over 2 \pi} {1 \over
\left(\phi^2 - f^2 \right)}. \label{eta}
\end{eqnarray}

Defining $\phi_e$ as the value of the field $\phi$ at the end of
inflation, $\epsilon\left(\phi_e\right) = 1$ we find two possible
solutions:
\begin{equation}
\left(\phi_e\right)^2 = f^2 + {M_{Pl}^2 \over 2 \pi} \left( 1 \pm
\sqrt{1 + 4 \pi f^2/M_{Pl}^2} \right).
\end{equation}
The solution with positive sign has $|\phi_e| > f$ and corresponds
to large-field inflation; that is, the field starts 
at $|\phi_i| > |\phi_e|> f$ 
and slowly rolls to values close to $f$ until inflation stops.
The solution with negative sign has $|\phi_e| < f$ and corresponds
to small-field inflation: the field starts near the local maximum
at $|\phi_i|<|\phi_e| <f $ and slowly rolls to values closer to $f$ until the end of
inflation.

The number $N$ of e-folds remaining until the end of inflation in
terms of the value of the field $\phi_N$ required to accomplish
this number of e-folds is given by:
\begin{equation}
N = {\sqrt{4 \pi} \over M_{Pl}} \int_{\phi_e}^{\phi_N} {d \phi
\over \sqrt{\epsilon(\phi)}} = {\pi \over M_{Pl}^2} \left(
\phi_N^2 - \phi_e^2 - f^2 \ln\left(\phi_N^2 / \phi_e^2 \right)
\right). \label{N}
\end{equation}
There is a bound on the maximum number of e-folds between the time that 
the present Hubble radius leaves the horizon 
and the end of inflation. The bound derives from limits on the 
gravitational wave background (provided that the
energy density of the Universe does not drop faster than that of radiation
after inflation) and is given roughly by $N_{\mbox{max}} \simeq 60$
\cite{DH,LL}. In the following we will use $N=50$ and $60$ for illustration
and denote the corresponding field value by $\phi_{N}$.

Once we solve Eq. (\ref{N}) for $\phi_{N}$, we can
immediately compute the spectral index of perturbations, $n_s$, and
the ratio of tensor to scalar perturbations, $r$ \cite{KKMR}:
\begin{eqnarray}
n_{s,N} &=& 1 - 4 \epsilon\left(\phi_{N}\right) + 2
\eta\left(\phi_{N}\right) \\
r_N &=& 16 \epsilon\left(\phi_{N}\right)
\end{eqnarray}

\begin{table}
\caption{Large-field Inflation: For different values of the vacuum expectation value $f$, we show 
the value of the scalar field ($\phi_{50}$, $\phi_{60}$) 
when the present Hubble radius crosses outside the horizon 
and at the end ($\phi_{e}$) of inflation, all in units of the Planck
mass. $n_N$ and $r_N$ denote the corresponding values of the scalar spectral 
index and the tensor-to-scalar ratio.}
\begin{ruledtabular}
\begin{tabular}{l|l|l|l|l|l|l|l}
$f$ & $\phi_e$ & $\phi_{50}$ & $\phi_{60}$ & $n_{50}$ & $n_{60}$  & $r_{50}$ & $r_{60}$  \\
\hline \hline
0.1 & 0.581  &4.04 & 4.41 & 0.941 & 0.951 & 0.313 & 0.262 \\
1.0 & 1.32  &4.48 & 4.84 & 0.946 & 0.955 & 0.280 & 0.237 \\
3.0 & 3.40  &6.17 & 6.49 & 0.954 & 0.961 & 0.229 & 0.195 \\
5.0 & 5.29  &8.06 & 8.37 & 0.956 & 0.963 & 0.207 & 0.176 \\
\end{tabular}
\end{ruledtabular}
\label{large}
\end{table}

\begin{figure}[htb]
\vspace*{-5mm}
\hspace*{-0.9cm}\includegraphics[height=18cm,width=13cm]{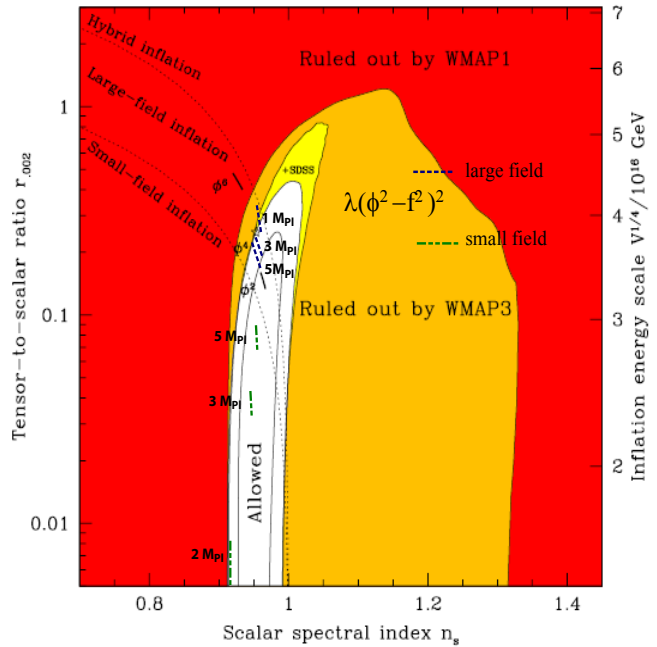}
\vspace*{-9cm}
\caption{Regions in the $r$ vs. $n_s$ plane excluded by WMAP1 (in red), WMAP3 (in beige), 
and by WMAP3+SDSS (in yellow), from \cite{Tegmarketal}. The two contours in the white 
region show those allowed at 68\% and 95\% CL. Regions occupied by chaotic inflation 
models with $\phi^2$, $\phi^4$, and $\phi^6$ potentials are indicated. We have 
superimposed regions occupied by our model in the large-field regime for 
$f=1, 3$, and $5 M_{Pl}$ (black, dashed) and in 
the small-field regime for $f=2, 3$, and $5 M_{Pl}$
(green, dot-dashed). Moving down these curves, the number of e-folds before the end of 
inflation that the Hubble radius expands outside the horizon varies from $N=50$ to 60 (except for the $\phi^4$ case, in which $N=64$ was used).}
\label{figureWMAP}
\end{figure}

In Table \ref{large}, we show the
resulting values for $n_{s,N}$ and $r_N$ for different values of
the symmetry breaking scale $f$, for the large-field case, $|\phi_i|> f$. For $f < M_{Pl}$, we
have $\phi >> f$ throughout inflation; in this limit, the potential of Eqn.(\ref{vphi}) 
is close to that of a 
pure $\lambda \phi^4$ theory, and the resulting values of $n_s$ and $r$ are similar to 
those of the $\phi^4$ model. For larger values of $f$, $\phi_e/f$ is close to unity; in this 
regime, one can expand the potential during inflation around $\phi = f$ and find that it is closer to 
quadratic than quartic. For $f \gsim 3 M_{Pl}$, Fig.\ref{figureWMAP} shows that the 
resulting values of $n_s$ and $r$ 
are consistent with the $68\%$ limits from WMAP3+SDSS LRG for values of $N$ approaching 60.

At 95\% CL, the parameter range $f >1 M_{Pl}$  is allowed by the current data.

%

Results for the case of small-field inflation, $|\phi_i|<f$, 
are shown in Table \ref{small}. In this case, 
one can expand the potential
around $\phi=0$, resulting in $V(\phi) \propto f^4 (1 - 2
\phi^2/f^2)$. For $f \lsim M_{Pl}$, $\phi_{60}$ is
exponentially smaller than $\phi_e$ \cite{KM}, which is unnatural, especially 
since quantum fluctuations impose a lower bound on the field amplitude. 
In fact, for $f < 0.8 M_{Pl}$, we find no solutions
to Eqn. (\ref{N}) in the small-field case. In this regime, for $\phi_{60} \ll f$,
one finds $n_s \simeq M_{Pl}^2/(\pi f^2)$. As $f$ increases, there is a 
transition at $f \simeq M_{Pl}$, where the scalar spectral index gets
substantialy closer to 1. As shown in Fig.\ref{figureWMAP}, consistency 
with WMAP3+SDSS at 68\% CL requires 
$f \gsim 3 M_{Pl}$.

From Tables \ref{large} and \ref{small} and Eqns.(\ref{eps},\ref{eta}), we see that 
$\epsilon(\phi_N), \eta(\phi_N) \ll 1$ for most of the cases studied, 
validating the use of the slow-roll approximation. 

The fact that the symmetry-breaking scale $f$ must be near $M_{Pl}$ is potentially appealing 
from the theoretical point of view, since the latter is a fundamental mass scale 
of gravitational origin. However, the observational constraint that $f$ must be 
several times larger than the Planck mass could raise concern about the validity 
of the semi-classical field theory approach and about the possibility of large gravitational 
corrections to the theory that could destroy the requisite flatness of the scalar 
field potential. In this context, we note that models 
with two \cite{twoaxions} or more \cite{moreaxions} axions have been proposed, 
in which a linear combination can result in an effective scale $f$ larger than 
$M_{Pl}$ while the fundamental mass scales in the theory are below the Planck mass.

\begin{table}
\caption{Small-field Inflation: For different values of the vacuum expectation value $f$, we show 
the value of the scalar field ($\phi_{50}$, $\phi_{60}$) 
when the present Hubble radius crosses outside the horizon 
and at the end ($\phi_{e}$) of inflation, all in units of the Planck
mass. $n_N$ and $r_N$ denote the corresponding values of the scalar spectral 
index and the tensor-to-scalar ratio.}
\begin{ruledtabular}
\begin{tabular}{l|l|l|l|l|l|l|l}
$f$ & $\phi_e$ &$\phi_{50}$  & $\phi_{60}$ & $n_{50}$ & $n_{60}$ & $r_{50}$& $r_{60}$  \\
\hline \hline
1.0 & 0.757 & $0.0002$  & $0.00004$ & 0.682 & 0.682 & 0.0  & 0.0 \\
2.0 & 1.74 & 0.163 & 0.110 & 0.918 & 0.919 & 0.00862 & 0.00385 \\
3.0 & 2.73 & 0.770 & 0.639 & 0.951 & 0.956 & 0.0428 & 0.0282 \\

5.0 & 4.73 & 2.49  & 2.29  & 0.961 & 0.967 & 0.0893  & 0.0686 \\
\end{tabular}
\end{ruledtabular}
\label{small}
\end{table}

\section{Thawing the quintessence field}

\begin{figure}[htb]
\vspace*{-1mm}
\includegraphics[height=6cm,width=9cm]{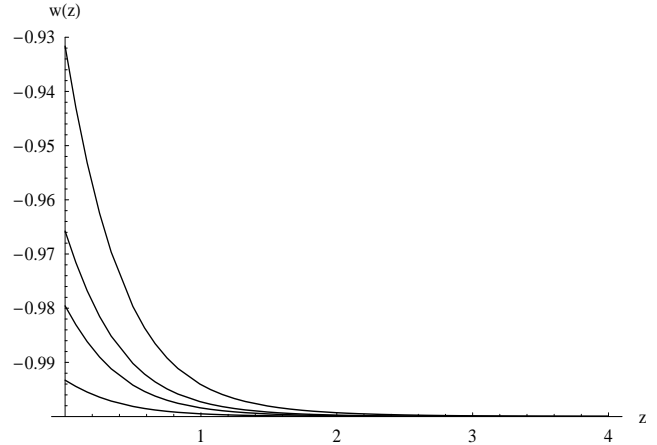}
 \caption{\label{wEvolution}
Evolution of the quintessence equation of state parameter $w(z)$ as a function of redshift for 
$f=5$ $M_{Pl}$, $\Omega_\varphi=0.7$, 
and $M^4 = 10 (5.4), 30 (3.2), 50 (2.5)$, and $100 (1.8)$ times $\rho^{(0)}_c$ 
(from bottom to top curves). 
The numbers in parentheses are the initial values of the field $\varphi_i$ in 
Planck mass units for the corresponding value of $M$. }
\end{figure}

It is interesting to study the consequences of this constraint 
on the symmetry breaking energy scale $f$ 
for the quintessence PNGB field in our model, described by the Lagrangian
\begin{equation}
{\cal L} = \frac{1}{2} (\partial_\mu \varphi)^2 - M^4 [ 1- \cos(\varphi/f) ].
\end{equation}
The quintessence behaviour is determined by three parameters, $f$, $M$, and the 
initial value $\varphi_i$ of the field when it was dynamically frozen in the early Universe. 
Once we fix values of two of the parameters, for instance $f$ and $M$, the value of $\varphi_i$ 
is determined by requiring that $\Omega_\varphi \simeq 0.7$ today. Since $\varphi$ only became 
dynamical at late times, the parameter $M^4$ must be of the order of the critical density 
today, cf. Eqn.(\ref{Mrho}), or larger.  
The steepness of the quintessence potential is measured by the mass of the 
$\varphi$ field, $m_\varphi \simeq M^2/f$. The larger the mass of the field, the earlier it can evolve, 
and therefore the larger the deviations of its equation of state from that of a cosmological 
constant, $w=-1$. We show this behaviour in figure \ref{wEvolution} by numerically solving the 
equations of motion for $\varphi$. 
We fix $f=5$ $M_{Pl}$, in keeping with the WMAP+SDSS constraints, and simultaneously vary $M$ and 
$\varphi_i$ to keep $\Omega_\varphi=0.7$ today, following \cite{quint}. 
We see that the PNGB quintessence equation of state $w(z)$ can evolve 
significantly at recent times for large values of $M^4$, making it distinct from a simple 
cosmological constant, as emphasized recently in the context of the so-called see-saw cosmology \cite{seesaw}.
For fixed $f$, the value of $M$ is bounded from below by the requirement that the quintessence field 
energy be large enough to dominate the Universe today and from above by the requirement that it drive 
accelerated rather than decelerated expansion. As pointed out in \cite{friemanetal,seesaw}, one 
can achieve evolution of the sort shown in Fig. \ref{wEvolution} without fine-tuning the mass 
parameters of the model. Such behavior is consistent with current constraints on the evolution of 
$w(z)$ (see, e.g., \cite{draganhiranya}) but could be tested by future projects aimed at probing 
the dark energy. 

%

\section{Conclusions}
In this Brief Report, we have studied the constraints on a simple model of quintessential 
inflation previously proposed by us \cite{us} that arise from the WMAP3 CMB and SDSS LRG data. 
We find that the effective scale of symmetry breaking, $f$, must be larger than about 3 $M_{Pl}$ 
in order to satisfy the constraints on the scalar spectral index and the tensor-to-scalar ratio 
from inflation. With these constraints, the resulting quintessence equation of state parameter $w(z)$ 
can nevertheless evolve rapidly at recent times, $z\lsim 1-2$, depending on the value of the induced
explicit symmetry breaking scale $M$, an example of a `thawing' dark energy model. Such models 
can be tested by precision probes of the dark energy equation of state expected over the coming 
decade.

\section*{Acknowledgments}

This work was partially supported by a CNPq-NSF binational
agreement, by the U.S. Department of Energy and NASA grant NAG5-10842 at Fermilab, and by 
the Kavli Institute for Cosmological Physics at the University of Chicago.
RR thanks Scott Dodelson at the Fermilab Particle Astrophysics Center and the 
Kavli Institute for Cosmological Physics for hospitality.

\end{document}